\documentclass[aps,pra,superscriptaddress,twocolumn]{revtex4}
\usepackage[utf8]{inputenc}
\usepackage{braket,textcomp}
\usepackage[usenames]{color}
\usepackage[dvipsnames]{xcolor}
\usepackage[breaklinks, colorlinks=true, urlcolor=blue,
anchorcolor=blue, citecolor=blue, filecolor=blue, linkcolor=blue,
menucolor=blue, linktocpage=true, pdfproducer=medialab,
pdfa=true]{hyperref}
\usepackage{amssymb,amsmath,bbm,mathcomp}
\usepackage{amsfonts,amsthm,subfigure,nccmath}
\usepackage{times}
\usepackage{mathrsfs}
\usepackage[mathscr]{euscript}
\usepackage{calligra}
\usepackage{graphicx,times}
\usepackage{epsfig}
\usepackage{cancel}
\usepackage{pifont}
\usepackage{enumerate}
\usepackage{mhsetup}
\usepackage{mathtools}
\usepackage{bm}
\usepackage{empheq}

\newcommand{\proj}[1]{\ket{#1}\!\bra{#1}}

\newcommand{\defi}{\coloneqq}
\newcommand{\dket}[1]{\ket{#1}\!\rangle}
\newcommand{\dbra}[1]{\langle\!\bra{#1}}
\newcommand{\smc}{{_{{C}\,}}}
\newcommand{\smg}{{_{{\Gamma}\,}}}

\newcommand{\exval}[1]{\langle{#1}\rangle}

\begin{document}
\title{A magnetic clock for a harmonic oscillator}
\author{Alessandro Coppo}
\address{Dipartimento di Fisica e Astronomia, Universit\`a di Firenze, I-50019,
Sesto Fiorentino (FI), Italy}
\address{INFN, Sezione di Firenze, I-50019, Sesto Fiorentino (FI), Italy}
\address{ISC-CNR, UOS Dipartimento di Fisica, Universit\`a ``La Sapienza'', I-00185, Rome, Italy}
\author{Alessandro Cuccoli}
\address{Dipartimento di Fisica e Astronomia, Universit\`a di Firenze, I-50019,
Sesto Fiorentino (FI), Italy}
\address{INFN, Sezione di Firenze, I-50019, Sesto Fiorentino (FI), Italy}
\author{Paola Verrucchi}
\address{ISC-CNR, UOS Dipartimento di Fisica, Universit\`a di Firenze, I-50019,
Sesto Fiorentino (FI), Italy}
\address{Dipartimento di Fisica e Astronomia, Universit\`a di Firenze, I-50019,
Sesto Fiorentino (FI), Italy}
\address{INFN, Sezione di Firenze, I-50019, Sesto Fiorentino (FI), Italy}

\begin{abstract} 
We present an implementation of a recently proposed 
procedure
for 
defining {\it time}, based on the description of the evolving 
system and its clock as non-interacting, entangled systems, according to the 
Page and Wootters approach.
We study how the quantum dynamics transforms into a classical-like 
behaviour when conditions related with macroscopicity are met by the 
clock alone, or by both the clock and the evolving system. 
In the description of this emerging behaviour finds its place the 
classical notion of time, as 
well as that of phase-space and trajectories on it. This allows us to 
analyze and discuss the relations that must hold between 
quantities that characterize system and clock separately, 
in order for the resulting overall picture be that of a physical 
dynamics as we mean it.
\end{abstract} 
\maketitle

\section{Introduction}
\label{s.introduction}

In the conventional formulation of Quantum Mechanics (QM) 
\cite{Dirac_PQM_1958, Peres2002} time is 
an external parameter, not related with any observable of the system;
this somehow anomalous status is often considered a weakness of the 
theory, referred to as "the problem of time" \cite{MugaME2002book},
mining the capacity of QM 
to describe the facets of the physical world.
One of the most promising proposal for "solving" such problem by 
treating time as a quantum observable emerged some decades 
ago\cite{PageW83} in the 
realm of quantum information,
and goes under the name of "Page and Wootters (PaW)
mechanism". The mechanism is based on the idea that the 
actual time $t$ for an evolving system is set by the fact that
another system (referred to as the "clock") be in a state labelled 
by the value $t$. The idea takes shape in the quantum formalism, with 
system and clock together assumed in an entangled state. The PaW 
proposal has been the subject of careful analysis in the following decades
and a convincing answer has been finally found for many criticisms originally raised against 
it \cite{GiovannettiLM15,MarlettoV17,BryanM18,SmithA19,AltaieHB22}. 
Moreover, in recent years it has been shown how the PaW mechanism can be 
employed to operationally define local time reference frames associated 
with several quantum clocks, allowing for time dilation and gravitational 
interaction\cite{Castro-RuizGBB20}, and a generalization of the PaW 
formalism has been proposed to investigate the causal structure of 
processes in general relativity\cite{BaumannKGB22}. The relationship 
of the PaW formalism with other approaches to quantum gravity and, 
more generally, to quantum mechanics\cite{Rovelli96} and to the 
definition of  (time-) reference frames, quantum observables and 
their classical limit has also been addressed\cite{Chataignier21}. 

Within the framework of the PaW mechanism, we have recently
\cite{FCBCVnat21} derived both the quantum Schr\"odinger equation 
and the classical Hamilton equations of motion
exclusively enforcing the conditions set by the PaW mechanism, 
in a way that consistently identifies the physical quantity that 
plays the role of time in both equations.

Aim of this work is to present an explicit realization of the procedure 
introduced in Ref.~\cite{FCBCVnat21}. 
Elements of our picture are two 
non-interacting and yet entangled systems: one is the clock $C$, 
described as a magnetic system, and the other is the evolving system 
$\Gamma$, chosen as a harmonic oscillator.
We will show that $C$ can describe the standard dynamics of $\Gamma$, 
regardless of whether the latter is quantum or classical. In addition, 
the explicit realization offers the possibility to carefully analyze all 
the details of the actual implementation of the PaW mechanism, allowing 
us to reveal the connection between the physical properties of $C$ 
and its capability of properly describe the dynamics of $\Gamma$. 


The structure of the paper is as follows: in Sec.~\ref{s.overall} we 
introduce the pair $\Gamma$ and $C$, with details about their respective 
Hamiltonians. In Sec.~\ref{s.QQ} we discuss the fully quantum model,
defining all the relevant quantities characterizing the entangled state 
of the two subsystems, while in Sec.~\ref{s.CQ}, after having introduced 
the proper formalism allowing us to let just one of the two subsystems
(the clock $C$) to become macroscopic and amenable of a classical description, 
we are able to explicitly relate the parameters describing $C$
with the energy of the system $\Gamma$ and its dynamics, showing how the energy
scale of $C$ reflects on its ability to describe the dynamics 
of the evolving system $\Gamma$ in its entire Hilbert space. In Sec.~\ref{s.CC} the
classical limit of both $C$ and $\Gamma$ is taken, in such a way to not 
only recover the classical Hamilton's Equations of Motion (EoM), but getting also 
directly the classical orbits in phase space. Finally, in the last Section,
we discuss the main results and draw our conclusions.


\section{The overall quantum system}
\label{s.overall}

In Quantum Mechanics (QM) any physical system is described by a theory
defined by some representation of a Lie-algebra $\mathfrak{g}$, 
i.e. by a Hilbert space ${\cal H}$ and the commutation relations 
$[\,\cdot,\cdot\,]$ between the operators acting on ${\cal H}$ that 
describe 
the observables of the system. 
The specific algebra is identified by requiring that 
the Hamiltonian $\hat{H}$ ruling the dynamical evolution of the system 
belongs to $\mathfrak{g}$.

In this work we consider two systems, $\Gamma$ and $C$.
\\
$\Gamma$ belongs to the family of {\it bosonic systems}, described by 
the Lie algebra $\mathfrak{h}_4$, 
whose representation on some infinite-dimensional Hilbert space is 
spanned by the operators
$\{\hat{a},\hat{a}^\dagger,\hat{n}=\hat{a}^\dagger\hat{a},\hat{\mathbb{I}}\}$ 
such that $[\hat{a},\hat{a}^\dagger]=\hat{\mathbb{I}}$.
In particular, $\Gamma$ is a harmonic oscillator with mass $M$ and 
frequency $\omega$. For reasons that will be clearer later, we introduce 
the  normalized bosonic operators
$\hat{\mathsf{a}}^{(\dagger)}:=\hat{a}^{(\dagger)}/\sqrt{M}$, such that
\begin{equation}
\left[\hat{\mathsf{a}},\hat{\mathsf{a}}^\dagger\right]=
\frac{\hat{\mathbb{I}}}{M}~,
\label{e.normalized_bosonic}
\end{equation}
in terms of which, setting $\hbar=1$ from now on, the Hamiltonian of $\Gamma$ reads
\begin{equation} 
\hat{H}_\smg:=\frac{\hat{p}^2}{2M}+\frac{1}{2}M\omega^2\hat{q}^2
=\omega M\left(\hat{\mathsf{n}}+\frac{\hat{\mathbb{I}}}{2M}\right)
\label{e.H_Gamma}
\end{equation}
with $\hat{\mathsf{n}}:=\hat{\mathsf{a}}^\dagger\hat{\mathsf{a}}$,
$\hat{q} 
:=\sqrt{\frac{1}{2\omega}}(\hat{\mathsf{a}}^\dagger+\hat{\mathsf{a}})$,
and $\hat{p} 
:=M\sqrt{\frac{\omega}{2}}(\hat{\mathsf{a}}^\dagger-\hat{\mathsf{a}})$.

In this work $\Gamma$ is the {\it evolving system}, whose time is 
possibly marked by $C$, as defined below.

$C$ belongs to the family of {\it spin systems}, described by 
the Lie algebra $\mathfrak{su}(2)$, 
whose representation on some $(2J+1)$-dimensional Hilbert space is 
spanned by the operators
$\{\hat{J}_0,\hat{J}_1,\hat{J}_2\}$ such that
$[\hat{J}_i,\hat{J}_j]=i\epsilon_{ijk}\hat{J}_k$ with $i(j,k)=0,1,2$, 
and the Casimir $\hat{J_0}^2$+$\hat{J_1}^2$+$\hat{J_2}^2$=
$J(J+1)\hat{\mathbb{I}}$ that provides the {\it spin} $J$ of the system.
In particular, $C$ is a spin-$J$ system in the presence of a 
magnetic field pointing in a direction labelled by the above index $0$. 
We use the normalized spin operators $\hat{\mathsf{j}}_i:=\hat{J}_i/J$, 
such that 
\begin{equation}
    [\hat{\mathsf{j}}_i,\hat{\mathsf{j}}_j]=
\frac{i}{J}\,\hat{\mathsf{j}}_k\,\epsilon_{ijk}\,,
\label{e.normalized commutators}
\end{equation}
in terms of which the Hamiltonian of $C$ reads
\begin{equation}
\hat{H}_\smc=\epsilon J\,\hat{\mathsf{j}}_0+F\,\hat{\mathbb{I}}~, 
\label{e.H_C}
\end{equation} 
with $\epsilon J$ and $F$ positive energies.
In this work $C$ is the {\it clock} that marks the time for the 
evolution of $\Gamma$. 

Referring to the PaW mechanism, we assume that $\Gamma$ and $C$ do not 
interact, and consequently write the Hamiltonian of the composite system 
$\Psi=C+\Gamma$ as
\begin{equation}
\label{e.Ham_PaW} 
\hat{H}=\hat{H}_\smc\otimes 
\hat{\mathbb{I}}_\smg-\hat{\mathbb{I}}_\smc\otimes\hat{H}_\smg~;
\end{equation} 
moreover, consistently with the fact that there is no other 
system, i.e. $\Psi$ is isolated, we 
require it to be in a pure state $\dket{\Psi}$ such that
\begin{eqnarray}  
&~&\hat{H}\dket{\Psi}=0~,\label{e.PaW_stationary}\\ 
&~&\dket{\Psi}\; \mbox{is entangled}~,
\label{e.PaW_entangled}
\end{eqnarray} 
where the double bracket in $\dket{~\cdot~}$ is used to indicate a 
vector in a Hilbert space that is the tensor product of two 
distinct Hilbert spaces or, which is the same, remind one that $\Psi$ 
is a composite system made of the two distinct subsystems $\Gamma$ and 
$C$.
The conditions \eqref{e.PaW_stationary}-\eqref{e.PaW_entangled} set the 
framework for the PaW mechanism and are quite often questioned, 
particularly as far as the possibility that an eigenstate of a 
non-interacting Hamiltonian be entangled. Although this should not 
surprise (just think to the maximally entangled Bell singlet of two qubits A and B, which is 
one of the eigenstates of a non-interacting Hamiltonian 
proportional to $\hat J^A_0+\hat J^B_0$), it is true that the PaW 
constraints are fulfilled only for some specific states of 
$\Gamma$ and $C$, whose identification is not trivial, as we show below.

\section{A quantum magnetic clock for a quantum oscillator}
\label{s.QQ}
In this section we use a quantum description for both $C$ and $\Gamma$. 
Their Hilbert spaces 
$\mathcal{H}_\smc$ and $\mathcal{H}_\smg$, spanned by
$\{\ket{J,m},~m=-J,-J+1,...,J\}$ and 
$\{\ket{n},~n\in\mathbb{N}\}$ respectively, are irreducible 
representations of $\mathfrak{su}(2)$ and $\mathfrak{h}_4$ defined via
\begin{equation}
\begin{cases}
\hat{\mathsf{j}}_0\ket{J,m}_\smc=\frac{m}{J}\ket{J,m}_\smc
\;\;\; 2J\in\mathbb{N},\;\; m=\{-J,...,J\}~,\\
\hat{\mathsf{n}}\ket{n}_\smg=\frac{n}{M}\ket{n}_\smg\;\;\; n\in\mathbb{N}~;
\end{cases}
\end{equation} 
without loss of generality we take $F=\epsilon J$ so that 
$\hat{H}_\smc\ket{J,-J}_\smc=0$. 

Any state of $\Psi$ can be written as %
\begin{equation}\label{e.global_state} 
\dket{\Psi}=
\sum_{n=0}^\infty\sum_{m=-J}^{J}c_{nm}\ket{J,m}_\smc\ket{n}_\smg\;\; 
\mbox{with}\;\;\sum_{n,m}|c_{nm}|^2=1~, 
\end{equation} 
referred to which the constraint \eqref{e.PaW_stationary} takes the form
\begin{equation} 
\sum_{n,m}c_{nm}\left[\epsilon (m+J)-M\omega
\left(n+\frac{1}{2}\right)\right]\ket{J,m}_\smc\ket{n}_\smg=0~, 
\end{equation} 
implying 
\begin{equation} 
c_{nm}=c_m \;\delta\left[\epsilon (m+J)-
\omega\left(n+\frac{1}{2}\right)\right]~.
\label{e.coefficients}
\end{equation} 
This means that in order for the coefficients 
$c_{nm}$ to be different from zero the quantum numbers $n$ and $m$ must 
satisfy 
\begin{equation}
n+\frac{1}{2}=\kappa r \left(m+J\right)~,
\label{e.constraint_QQ} 
\end{equation}
with
\begin{equation}
\kappa:=\frac{\epsilon J}{\omega M}~{\rm and}~r:=\frac{M}{J}~.
\label{e.Kappa_r_def}
\end{equation} 
So, for the constraint \eqref{e.PaW_stationary} to hold, 
once $\kappa$ and $r$ are fixed, only 
some states $\ket{J,m}_\smc\ket{n}_\smg$ can enter the 
decomposition \eqref{e.global_state}. If we consider for instance 
$\kappa r=3/4$, 
it is $n+1/2=3(m+J)/4$, and hence, depending on the value of $J$, it is
\begin{eqnarray}
J<1 &\rightarrow&\nexists\, (m,n) \nonumber\\
J=1 &\rightarrow& (m=1,n=1)\nonumber\\
J=3/2 &\rightarrow& (m=1/2,n=1)\nonumber\\
J=2 &\rightarrow& (m=0,n=1)\nonumber\\
J=5/2 &\rightarrow& (m=-1/2,n=1)\nonumber\\
J=3   &\rightarrow& (m=-1,n=1)\, \vee \,(m=3,n=4)\nonumber \\
... ~,
\label{e.somecases} 
\end{eqnarray}
where $\rightarrow$ stands for "quantum numbers of the allowed states".
Moreover, since $\dket{\Psi}$ must be 
entangled, as required by \eqref{e.PaW_entangled}, 
there must be at least two pairs 
$(m,n)$ that satisfy Eq.~\eqref{e.constraint_QQ}; in the above 
considered case $\kappa r=3/4$, for instance, it must hence be $J\geq 
3$. 
In the most general case it can be shown (see App. \hyperlink{a.A}{A}) that there 
exist pairs $(m,n)$ such that conditions \eqref{e.PaW_stationary} and
\eqref{e.PaW_entangled} hold for $\dket{\Psi}$ if 
\begin{eqnarray}
\kappa r&=&\frac{2i_n+1}{2i_m}~~,~~i_m,i_n\in{\mathbb{N}}~,\label{e.Kappa_constr}\\
m&=&i_m(2l+1)-J~~,~~ n=i_n(2l+1)+l~,\label{e.mn_constr}\\
&~&{\rm with}~l=0,1,2,...,\left\lfloor J/i_m-1/2\right\rfloor~,\label{e.def_elle}\\
&~&{\rm and}~J\ge 3 i_m/2~,\label{e.J_constr}
\end{eqnarray}
where the last inequality ensures that there are at least two 
allowed $(m,n)$ pairs, i.e. that $\dket{\Psi}$ is entangled.
In the previous example $\kappa r=3/4$, it is $i_n=1,i_m=2$, $J\geq 3$, and 
$\{(m,n)\}=\{(4l+2-J,3l+1),\,l=0,1,..,\lfloor 
J/2-1/2\rfloor\}$.

When Eq.~\eqref{e.constraint_QQ} holds, the coefficients in 
\eqref{e.global_state} depend on one index only, say $m$, and one can 
write
\begin{equation}\label{e.para_Dicke}
\dket{\Psi}=\sum_{m\in \mathcal{A}} c_m\ket{J,m}_\smc\ket{n_m}_\smg~,  
~n_m\defi\kappa r (m+J)-\frac{1}{2}~,
\end{equation}
with $\mathcal{A}$ the set of integers $m$ consistent with 
Eqs.~\eqref{e.Kappa_constr}-\eqref{e.J_constr}.
Following Ref.~\cite{FCBCVnat21}, we introduce Generalized Coherent 
States (GCS) for the clock, i.e. the $\mathfrak{su}(2)$-CS also known 
as Spin Coherent States (SCS), defined as
\begin{equation}
\ket{\Omega}_\smc
:= e^{\Omega \hat{J}_+-\Omega^* \hat{J}_-}
\ket{J,-J}_\smc~,~~
\ket{\Omega}\leftrightarrow \Omega(\theta,\varphi)\in S^2~,
\end{equation}
where $\hat{J}_\pm=\hat{J}_1\pm i\hat{J}_2$, 
and $\Omega(\theta,\varphi)=(\theta/2)e^{-i\varphi}$.
These states are in one-to-one correspondence with the points of the 
sphere $S^2$, as easily seen by recognizing 
$(\theta,\varphi)\in[0,\pi]\times[0,2\pi)$ as the usual polar 
coordinates. 

Partially projecting $\dket{\Psi}$ upon SCS of $C$,  
the following unnormalized elements of $\mathcal{H}_\smg$ are obtained 
\begin{eqnarray}
    &~&\ket{\Phi_\theta(\varphi)}_\smg
\defi\braket{\Omega|\Psi}\!\rangle=
\sum_{m\in \mathcal{A}}c_m\braket{\Omega|J,m}\,\ket{n_m}_\smg\nonumber\\
    &~&=\sum_{m\in \mathcal{A}}c_m\binom{2J}{m+J}^{1/2}
\left(\cos\frac{\theta}{2}\right)^{J-m}
\left(\sin\frac{\theta}{2}\right)^{J+m}\nonumber\\
    &~&\qquad\qquad\quad e^{-i\varphi(J+m)}\ket{n_m}_\smg~,
\label{e.unnormalized_Phi}
\end{eqnarray}
where the specific form of $\braket{\Omega|J,m}$ can be found, for 
instance, 
in Ref.~\cite{Perelomov86}. Notice that taking $\varphi$ in $[0,\infty)$, 
rather than in $[0,2\pi)$, implies no ambiguity and can be safely done.

The state $\ket{\Phi_\theta(\varphi)}_\smg$
obeys~\cite{FCBCVnat21}
\begin{equation}\label{e.un-Schroedinger}
i\epsilon\frac{d}{d\varphi}\ket{\Phi_\theta(\varphi)}_\smg=
\hat{H}_\smg\ket{\Phi_\theta(\varphi)}_\smg~,
\end{equation}
a differential equation that has the form of the 
Schr\"odinger equation, with $\varphi/\epsilon$ as time, and 
yet does not describe the 
unitary dynamics of pure states, for two reasons: first, 
$\ket{\Phi_\theta(\varphi)}_\smg$ has a 
dependence on a further external parameter $\theta$ that 
makes no sense if $\Gamma$ is isolated, as required for the 
Schr\"odinger equation to hold.
Second, $\ket{\Phi_\theta(\varphi)}_\smg$ is not 
a physical state of $\Gamma$ because it is unnormalized.
In fact, this latter reason is most often considered 
amendable (as done for instance in Refs.~\cite{PageW83,GiovannettiLM15}), since the same differential equation 
\begin{equation}\label{e.Schroedinger}
i\epsilon\frac{d}{d\varphi}
\ket{\phi_\theta(\varphi)}_\smg=
\hat{H}_\Gamma\ket{\phi_\theta(\varphi)}_\smg~,
\end{equation} 
holds for the normalized state
\begin{equation}
\ket{\phi_\theta(\varphi)}_\smg:=
\frac{\ket{\Phi_\theta(\varphi)}_\smg}{\chi^2(\theta)}~,
\label{e.normalized_phi}
\end{equation}
given that 
\begin{eqnarray}
&~&\chi^2(\theta):=
\exval{\Phi_\theta(\varphi)|\Phi_\theta(\varphi)}_\smg=
\sum_{m\in \mathcal{A}}|c_m|^2 \,\left|\braket{\Omega|J,m}\right|^2 
\label{e.chi2_def}\\
    &~&=\sum_{m\in \mathcal{A}}|c_m|^2 
\binom{2J}{m+J}\left(\cos\frac{\theta}{2}\right)^{2(J-m)} 
\left(\sin\frac{\theta}{2}\right)^{2(J+m)}~,\label{e.chi2} 
\end{eqnarray} 
is a positive function that does not depend on $\varphi$.
Notice that Eq.~\eqref{e.normalized_phi} defines a physical state for 
$\Gamma$ for any $\theta$ such that $\chi(\theta)\neq 0$.
In fact, the role played by $\chi^2(\theta)$ goes well beyond its being the 
norm of $\ket{\Phi_\theta(\varphi)}_\smg$, as will be clear once the 
parametric 
representation with GCS is introduced.
Whether one thinks the above {\it ad hoc} normalization to be a 
satisfactory solution or not, neither Eq.~\eqref{e.un-Schroedinger} nor its 
sibling for the normalized state $\ket{\phi_\theta(\varphi)}_\smg$ 
should be considered equations of motion, at this stage, as best 
explained by the following example.

\begin{figure} \includegraphics[scale=0.45]{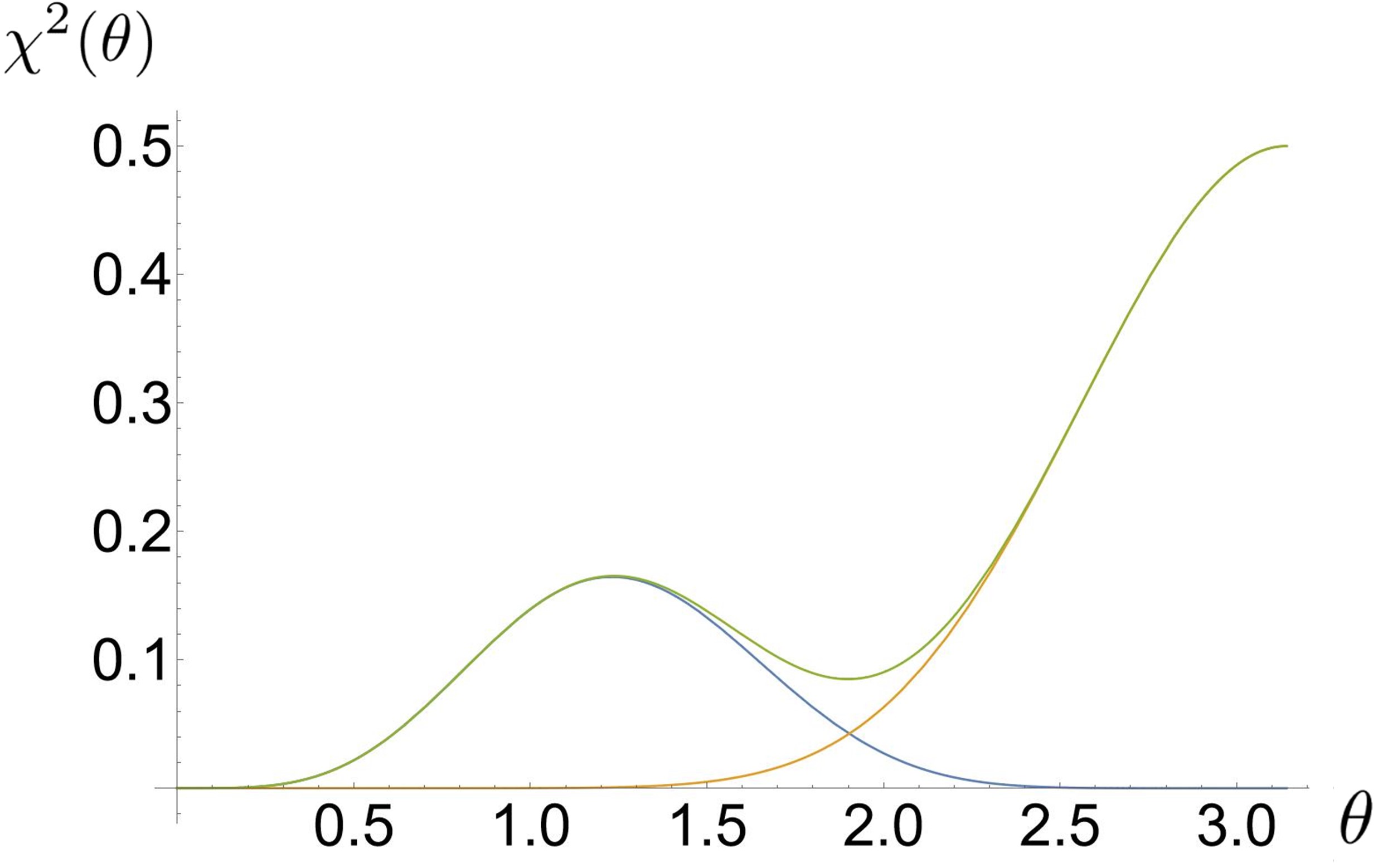} 
\caption 
{Graphical representations of $\frac{1}{2}\left|\braket{\Omega|3,-1}\right|^2$ (blue line), $\frac{1}{2}\left|\braket{\Omega|3,3}\right|^2$ (orange line), and $\chi^2(\theta)$ (green line), as defined in Eqs.\eqref{e.chi2}-\eqref{e.chi2_J.3}.}
\label{f.1}
\end{figure}

Let us take $\kappa r=3/4$ and $J=3$ in Eq.~\eqref{e.constraint_QQ}, 
i.e. (see \eqref{e.somecases})
$c_{-1}=c_3=1/\sqrt{2}$, and $c_m=0$ otherwise.
In this case, the sum in Eq.~\eqref{e.chi2} reduces to just two terms, 
\begin{equation}
 \chi^2(\theta)=\frac{15}{2} 
\left(\cos\frac{\theta}{2}\right)^8
\left(\sin\frac{\theta}{2}\right)^4+
\frac{1}{2}\left(\sin\frac{\theta}{2}\right)^{12}~,
\label{e.chi2_J.3}
\end{equation}
that simultaneously vanish
for $\theta=0$ only, as seen in Fig.~\ref{f.1}. Therefore, 
any  
$\theta\in(0,\pi)$ defines a normalized state for $\Gamma$, that reads
\begin{eqnarray}
 \ket{\phi_\theta(\varphi)}_\smg&=&
\frac{\sqrt{15}\left(\cos\frac{\theta}{2}\right)^4}
{\sqrt{15\left(\cos\frac{\theta}{2}\right)^8+
\left(\sin\frac{\theta}{2}\right)^8}}e^{-2i\varphi}\ket{1}_\smg\nonumber\\
 &+&\frac{\left(\sin\frac{\theta}{2}\right)^4}
{\sqrt{15\left(\cos\frac{\theta}{2}\right)^8+
\left(\sin\frac{\theta}{2}\right)^8}}e^{-6i\varphi}\ket{4}_\smg~.
\label{e.phi_specific-case}
\end{eqnarray}
This expression shows one of the most relevant features of this 
fully quantum setting, namely that once the 
parameters of the Hamiltonians \eqref{e.H_Gamma} and \eqref{e.H_C},
are fixed, the system $C$ can mark the time for the evolution of 
$\Gamma$ via the PaW mechanism only if its dynamics is limited to a 
subspace of $\mathcal{H}_\smg$, defined by $M,\omega,\epsilon$, and the 
Casimir $J$. In the above example, where $\kappa r=3/4$ and $J=3$, for 
instance, 
the $\varphi$-derivation of the state \eqref{e.phi_specific-case} 
leads to a differential equation that cannot describe the appearance of 
elements of ${\cal H}_\smg$ others than $\ket{1}_\smg$ and 
$\ket{4}_\smg$.
In general, the finiteness of the Casimir $J$ in 
Eqs.~\eqref{e.mn_constr} and \eqref{e.def_elle} 
sets an upper limit for $n$, which prevents
$\ket{\phi_\theta(\varphi)}_\smg$ to 
explore its infinite-dimensional Hilbert space as time goes by.
One could clearly observe that it is absolutely not surprising that
difficulties arise when one tries to label states of an infinite-dimensional 
Hilbert space by states of a finite dimensional one. However, we see
that even if we consider the equally possible symmetrical setting, where 
the harmonic oscillator plays the role of the clock and the spin system
is the evolving system whose dynamics we are interested to describe, we 
can still be unable to properly explore the full Hilbert space of the 
evolving system, unless further adjustment is made, e.g. properly shifting
the ground state of the clock or the evolving system.

%
This result may sound puzzling, as we are used to consider clocks as 
objects whose capability of marking the time does not depend on the 
specific evolution that unfolds in such time, no matter whether in 
quantum or classical physics. A useful analogy to solve this conundrum 
comes from considering any model where an external magnetic field {\it 
is applied} to a quantum magnetic system via a Zeeman interaction
${\mathbf h}\cdot{\hat{\mathbf{S}}}$. In this case, the field is by all 
means a classical vector, introduced "by hand" according to some 
phenomenological evidence, and never related to a second quantum system 
entering the scene together with the one to which the field is applied, 
described by the three spin operators $\hat{S^x},\hat{S^y},\hat{S^z}$. 
However, it can be naively understood, but also formally 
demonstrated~\cite{2017RFCTVP}, that a Zeeman term effectively 
describes a Heisenberg-like interaction between two quantum magnetic 
systems, $\hat{\mathbf h}\cdot{\hat{\mathbf{S}}}$, one of which has such 
a large value of the Casimir $h$ that not only the dimension of its 
Hilbert space can be considered infinite but, most importantly, 
the observable associated to the quantum operator $\hat{\mathbf h}$ 
can be described as a classical vector ${\mathbf h}$~\cite{2020CCFV}, 
continuously defined on a sphere of radius $h$.  


With this analogy in mind, and 
based on the results of Ref.~\cite{FCBCVnat21} we claim that in order to get a 
proper setting, where time has the same status that we give it in 
standard QM and classical physics, the clock must be effectively 
described as a classical system.
As for the dependence of 
$\ket{\Phi_\theta(\varphi)}_\smg$ on $\theta$, in Ref.~\cite{FCBCVnat21} 
it is shown that clues about its meaning emerge already in this full 
quantum setting. However, as things get clearer when 
the quantum-to-classical crossover for $C$ is taken, we move the 
discussion to the next Section, where we use a hybrid scheme
where the description stays quantum for $\Gamma$ and becomes classical 
for $C$.

\section{A classical magnetic clock for a quantum oscillator}
\label{s.CQ}
In this section we want to keep describing $\Gamma$ by QM, while
introducing the formalism of classical physics for $C$, and $C$ only. 
This implies a fundamental distinction between the two systems, whose 
analysis requires some 
special tools. In fact, when dealing with composite systems in entangled 
states, the 
usual way to focus upon one component only is by partial tracing on 
the Hilbert space of the uninteresting part to get the density operator,  
which is the main tool for Open Quantum Systems (OQS) analysis. 
However, another strategy can be adopted, 
such that the OQS state is represented by an ensemble of normalized 
elements of its Hilbert space, i.e. pure states, depending on 
parameters that refer to the uninteresting part, a dependence 
exclusively due to the entanglement between components.
This strategy leads to so called {\it parametric representations},
that differ from each other in the parameters that characterize the ensemble.
Specific parametric representations are used for instance in 
Refs.~\cite{BornO27,Hunter75}, while a more general definition can be found in Ref.~\cite{2013CCGV},
where it is shown that the ensemble is always obtained by partially projecting 
the global state upon sets of normalized states of the 
uninteresting part, whose choice selects the all important parameters.
Amongst parametric representations, the one that better fits situations 
where the uninteresting system undergoes a quantum-to-classical 
crossover is the one defined by choosing the above states as 
GCS~\cite{2013CCGV}. 
When the GCS are the spin coherent states introduced above, the 
representation follows from writing $\dket{\Psi}$ in the form
\begin{equation}
\dket{\Psi}=\int_{S^2}d\mu(\Omega)\chi(\theta)\ket{\Omega}_\smc
\ket{\phi_\theta(\varphi)}_\smg~,
\label{e.para_Psi}
\end{equation}
as obtained by inserting 
$\int_{S^2}
d\mu (\Omega)\proj{\Omega}=\hat{\mathbf{I}}$ in Eq.~\eqref{e.global_state},
with $ d\mu(\Omega)=\frac{2J+1}{4\pi}\sin{\theta}d\theta d\varphi$
the measure on $S^2$.
Notice that $\ket{\phi_\theta(\varphi)}_\smg$ and $\chi(\theta)$ are 
the same as in Eqs.~\eqref{e.normalized_phi} and \eqref{e.chi2_def}, 
which helps understanding their meaning.
In fact, from 
\eqref{e.para_Psi} it follows\cite{2015LCV,2015LCV_ijtp} that the density operator
for the system $\Gamma$ is $\rho_\smg={\rm Tr}_{_C}\left(\dket{\Psi}\dbra{\Psi}\right)=
\int_{S^2}d\mu(\Omega)\chi^2(\theta)\proj{\phi_\theta(\varphi)}$ from which we understand $\chi^2(\theta)$ as the 
probability distribution on the parameter space that $\Gamma$ be in the 
state $\ket{\phi_\theta(\varphi)}_\smg$, which is the same~\cite{2015LCV} as 
the probability 
for $C$ to be in $\ket{\Omega}$, when $\Psi$ is in $\dket{\Psi}$.
Consistently with 
$\chi^2(\theta)$ being a probability distribution on $S^2$ it is  
$\int_{S^2}d\mu(\Omega)\chi(\theta)^2=1$.

As already mentioned, we want to enforce a classical description for the 
clock, a step that is not necessarily possible to take. In fact,
there is a formal procedure to check whether or not a quantum theory can 
flow into a classical one when the system that it describes becomes 
macroscopic. The procedure strongly relies on the properties of GCS 
that enter the derivation of 
Eq.~\eqref{e.un-Schroedinger}, and 
provides a general description of the so called quantum-to-classical 
crossover. Without entering into the details of this procedure, which is 
extensively treated in Refs.~\cite{Yaffe82,2013CCGV,2020CCFV,Coppo22},
we here assume that the quantum theory that describes $C$ flows into a 
well defined classical theory when $C$ becomes macroscopic, i.e. when 
$J\to\infty$ and the commutators \eqref{e.normalized commutators} 
vanish. Moreover, one of the rule dictated by the above mentioned 
procedure requires that the Hamiltonian operator $\hat{H}_\smc$ goes 
through the classical limit in such a way that its expectation values on 
GCS correspond to the values of a well-behaved function on the proper 
phase-space.
Since such expectation values read
\begin{equation}
\exval{\Omega|\hat H_\smc|\Omega}=
J\epsilon(1-\cos\theta)\defi E(\theta)~,
\label{e.Etheta}
\end{equation}
we require
\begin{equation}
\lim_{J\to\infty}J\epsilon<\infty~.
\label{e.JepsilonCQ}
\end{equation}
From Eq.~\eqref{e.Etheta} we also get a clue about the meaning of the 
somehow baffling parameter $\theta$: in fact, it can be 
demonstrated~\cite{2013CCGV} that 
\begin{equation}
\label{e.chi2_QC}
\chi^2(\theta)\underset{J\to\infty}{\longrightarrow}
\sum_{m\in 
\mathcal{A}}|c_m|^2\delta\left(m+J\cos\theta\right)~,
\end{equation}
where $\delta(~\cdot~)$ is the Dirac-$\delta$ distribution,
meaning that 
\begin{equation}
    \ket{\phi_\theta(\varphi)}_\smg
\underset{J\to\infty}{\longrightarrow}
e^{-i\varphi\, J(1-\cos\theta)}\,\ket{n_\theta}_\smg~,
\label{e.gammastate_CQ}
\end{equation}
with $n_\theta\defi n_{m=-J\cos\theta}$. 
Now, it is easily checked that condition \eqref{e.PaW_stationary} and 
Eq.~\eqref{e.gammastate_CQ} 
together make
\begin{equation}\label{e.stat_Schroedinger}
\hat{H}_\smg\ket{\phi_\theta(\varphi)}_\smg=
E(\theta)\ket{\phi_\theta(\varphi)}_\smg~,
\end{equation}
which is the stationary Schr\"odinger equation for $\Gamma$, with 
$E(\theta)$ as in \eqref{e.Etheta}.
This means that $C$ provides two parameters: 
$\varphi$, to describe the dynamics of $\Gamma$, and $\theta$ 
to set its energy; such conclusion well integrates into the remarks 
made in Ref.\cite{FCBCVnat21} about the role played by the clock's algebra,
PaW's constraint and entanglement from the perspective of understanding the
origin of the time-energy uncertainty relation for the evolving system $\Gamma$.

%
%

\begin{figure}
\includegraphics [width=\linewidth]{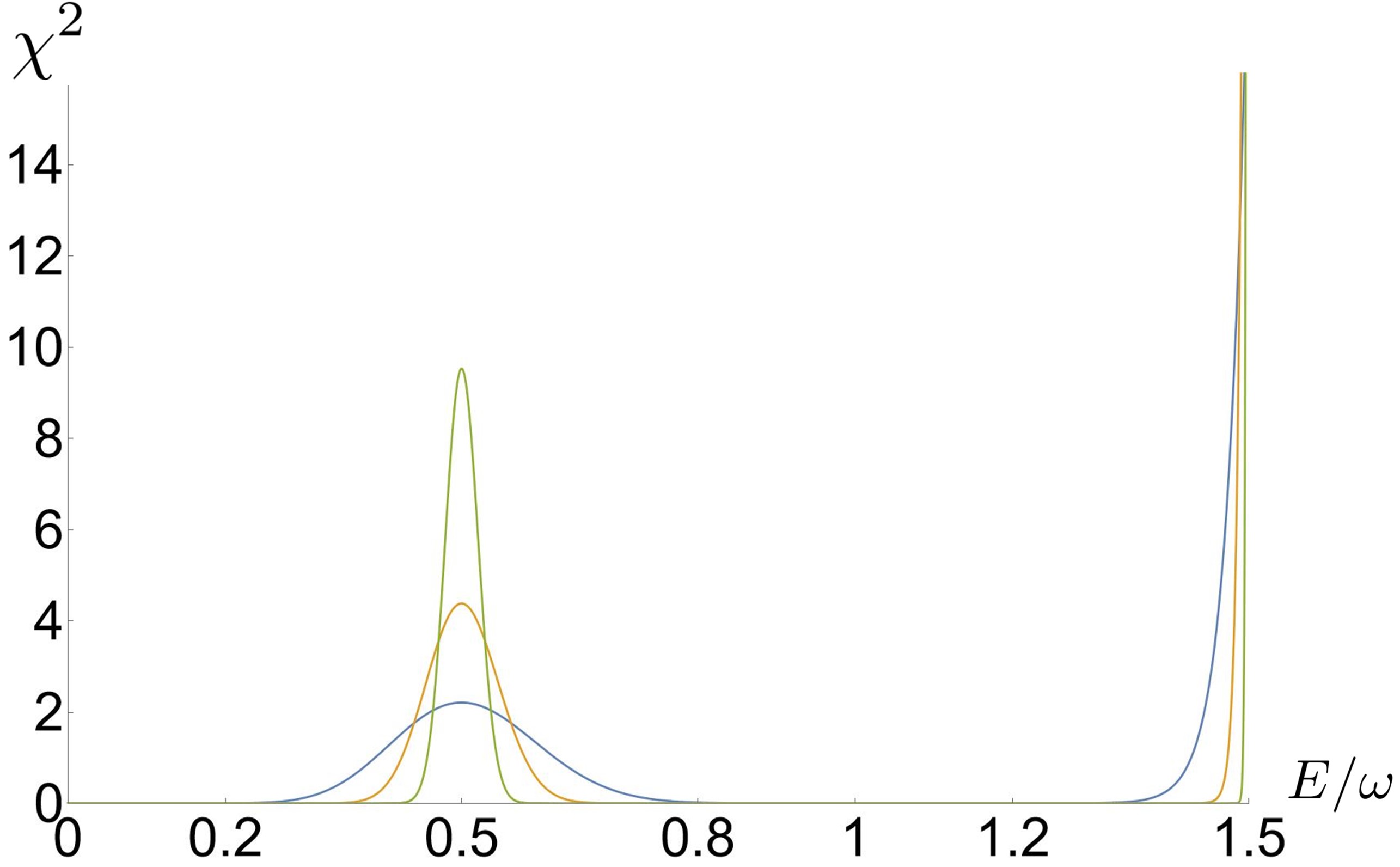}
\caption{Graphical representation of $\chi^2$ when $\kappa r J=3/4$ for $J=30$ (blue line), $J=120$ (orange line) and $J=570$ (green line).}
\label{f.2}
\end{figure}

If this result solve one of the puzzle of the above section, namely the 
physical meaning of the parameter $\theta$, we still have to check 
whether or not treating $C$ as a classical clock makes 
Eq.~\eqref{e.Schroedinger} capable of moving $\Gamma$ around in its 
entire Hilbert space.
To this aim, we get back to 
Eqs.~\eqref{e.Kappa_constr}-\eqref{e.J_constr} and notice that, 
despite the ratio $r$ vanishes in the $J\to\infty$ limit, the constraint 
\eqref{e.constraint_QQ} stays meaningful as far as $m=
\mu J$, with 
$\mu=(h-J)/J$ and $h=0,1,...2J$,
in which case it becomes
\begin{equation}
n+\frac{1}{2}=\kappa r J (\mu+1)~,
\label{e.constraint_CQ}
\end{equation}
and $\kappa r J= \frac{\epsilon J}{\omega}$ stays finite for 
$J\to\infty$ since we have enforced condition \eqref{e.JepsilonCQ}.
Therefore, with a bit of care, we can still refer to 
Eqs.\eqref{e.Kappa_constr}-\eqref{e.J_constr}.
Let us for instance take $\kappa r J = 3/4$: combining 
Eqs.~\eqref{e.J_constr} and \eqref{e.Kappa_constr} we find 
$i_n=0$, and $i_m=2J/3$. Therefore it is $l=0,1$, and the allowed 
pairs of quantum numbers are $m=-J/3,n=0$ and $m=J,n=1$. Consistently, according to Eq.~\eqref{e.chi2_QC}, the support of the function $\chi^2(\theta)$ tends to the set $\{\arccos{1/3},\pi\}$, that corresponds, via Eq.~\eqref{e.Etheta}, to the two admitted values $E=\omega/2$ and $E=3\omega/2$ of energy for $\Gamma$, as one can see from Fig.~\ref{f.2}. So, at first glance, it does not seem that taking the classical limit for $C$ improves the capability of $\Gamma$ to explore its Hilbert space via the usual Schr\"odinger equation. \\
On the other hand, we notice that taking 
$\kappa r J = 3/4$ while considering a very large $J$ implies 
$\kappa r \ll 1$, which is not a necessary condition. In fact, taking $\kappa r J \gg 1$ enables $\kappa r$ to stay finite when $J$ grows, and
to satisfy $1/(2\kappa r) = i_m \in \mathbb{N}$ with $i_m < \infty$ implying $i_n = 0$, so that, 
according to Eqs. \eqref{e.Kappa_constr}-\eqref{e.J_constr}, the quantum number $n$ 
can start from zero and grow larger and larger in order to set $\Gamma$
free to wander in $\mathcal{H}_\smg$. Moreover, whereas the condition $1/(2\kappa r) \in \mathbb{N}$ 
is essentially due to the nature of the algebras defining $\Gamma$ and $C$, the requirement 
 $\kappa r J \gg 1$ is very meaningful, as it means $\epsilon J \gg\omega$, that is to say 
that, in order for $C$ to properly work as a clock for $\Gamma$,
it must be big, amenable to a classical description, and characterized by a much bigger
energy-scale than that of the evolving system. We will further comment upon this point 
in the Conclusions. Before of that, in the next section, we conclude the
construction leading to the identification of time and consider what happens when also the
harmonic oscillator $\Gamma$ undergoes a quantum-to-classical crossover.


%
%

%
\section{A classical magnetic clock for a classical oscillator}
\label{s.CC}

In this last section we want to see if the above scheme also works in a 
completely classical setting, i.e. where we might find 
connections with general relativity and gravity.
To this aim we assume that $\Gamma$ is macroscopic and the theory that 
describes it fulfills the conditions ensuring the possibility of an 
effectively equivalent classical description. This is done as in 
Sec.~\ref{s.CQ} by introducing GCS for $\Gamma$, the usual 
Glauber coherent states for the harmonic oscillator (HCS), defined as 
\begin{equation}
\ket{\alpha}_\smg= 
e^{\alpha \hat{\mathsf{a}}-\alpha^* \hat{\mathsf{a}}^\dagger}\ket{0}_\smg 
\mbox{\;with\;} \ket{\alpha}\leftrightarrow \alpha \in \mathbb{C}~,
\end{equation}
where $\mathbb{C}$ is the complex plane. 
According to Eq.~\eqref{e.normalized_bosonic} the quantum-to-classical 
crossover is obtained for $M\to\infty$; moreover, as in the case of the 
magnetic system, the expectation values of $\hat 
H_\smg$ upon HCS
\begin{equation}
\braket{\alpha|\hat{H}_\smg|\alpha}= 
\omega \left(M|\alpha|^2+\frac{1}{2}\right)~:=E_\smg(|\alpha|)
\label{e.H_alpha}
\end{equation} 
must stay finite as $M\to\infty$, and we hence require
\begin{equation}
\lim_{M\to\infty} M\omega<\infty~.
\label{e.Momega_CC}
\end{equation}
Despite Eq.~\eqref{e.constraint_QQ} remains well defined in this limit, 
provided Eq.~\eqref{e.JepsilonCQ} holds, we notice that quantum numbers, 
let alone conditions upon them, should have no place in a completely 
classical description, suggesting that a different analysis must be used 
to get information about the classical configurations of $C$ and 
$\Gamma$ that are compatible with the assumptions of the PaW mechanism.
At first sight, it might seem impossible to translate conditions set in 
a genuinely quantum formalism into something which is meaningful also 
in classical physics. However,
one of the most relevant bonuses provided by parametric 
representations with GCS is that they allow us to move into the 
classical realm while keeping contact with the original quantum 
description. In fact, using the resolution of the identity upon ${\cal H}_\smg$ 
provided by the HCS, we write Eq.~\eqref{e.para_Psi} as
\begin{equation}\label{e. double para}
    \dket{\Psi}=
\int_{S^2}d\mu(\Omega)\int_{\mathbb{C}}d\mu(\alpha)\;
\beta(\Omega,\alpha)\ket{\Omega}_\smc \ket{\alpha}_\smg~,
\end{equation}
where $d\mu(\alpha)=\frac{M}{2\pi} d\alpha d\alpha^*$ is the measure 
on $\mathbb{C}$, and the square modulus of the function
\begin{equation}\label{e. beta 1}
  \beta(\Omega,\alpha):=\big(\,_{\smc\!\!\!}\bra{\Omega}\otimes\,_{\smg\!\!\!}\bra{\alpha}\big)\ket{\Psi}\rangle  
\end{equation}
is the conditional probability for $\Gamma$ to be in the state 
$\ket{\alpha}$ when $C$ is in the state $\ket{\Omega}$, or vice-versa,
given that the global system $\Psi$ is in $\dket{\Psi}$. Requiring that 
this probability be finite has the same meaning of Eq.~\eqref{e.constraint_QQ}.
In fact, only the classical configurations of clock and evolving 
system, defined by the respective phase-space coordinates $\Omega\in 
S^2$ and $\alpha\in\mathbb{C}$, and such that 
\begin{equation}
|\beta(\Omega,\alpha)|^2 > 0~,
\label{e.constraint_beta}
\end{equation}
are allowed configurations $(\Omega,\alpha)$ of the global system 
$\Psi$.
Using results from, for instance, Ref.~\cite{Perelomov86},  
reminding Eq.~\eqref{e.para_Dicke}, 
$n_m:=\kappa r(m+J)-\frac{1}{2}$, 
and $\Omega=\frac{\theta}{2}\exp^{-i\varphi}$, it is
\begin{align}
&\beta(\Omega;\alpha)=\sum_{m\in \mathcal{A}} 
c_m\braket{\Omega|J,m}\braket{\alpha|n_m} \label{e. beta 2}\\
&=\sum_{m\in \mathcal{A}} c_m \binom{2J}{m+J}^{1/2}\left(\cos\frac{\theta}{2}\right)^{J-m}\left(\sin\frac{\theta}{2}\right)^{J+m}\cdot\nonumber\\
&\qquad\quad \cdot\; e^{-i\varphi(J+m)}\quad 
e^{-\frac{M|\alpha|^2}{2}}\frac{(\sqrt{M} \alpha)^{n_m}}{\sqrt{n_m!}}\nonumber\\
 &\!\!\!\!\!\!\!\!\!\!\!\!\!\!\xrightarrow{J,M\rightarrow 
\infty}\sum_{m\in \mathcal{A}} c_m\;\; 
\delta\left(m+J\cos\theta\right)\,e^{-i\varphi(J+m)}\quad\cdot\nonumber\\
&\qquad\qquad\quad\cdot\,\delta\left(M|\alpha|^2-n_m\right)\;e^{-in_m\arg{\alpha}}~,
\label{e.beta 3}
\end{align}
and the condition \eqref{e.constraint_beta} translates into
\begin{equation*}
\begin{cases}
m=-J\cos\theta \\
M|\alpha|^2=n_m\sim\kappa r(m+J)~,
\end{cases}
\end{equation*}
where we have neglected $\frac{1}{2}$ w.r.t. $M|\alpha|^2$, for 
$M\to\infty$, i.e
\begin{equation}
\omega M|\alpha|^2=\epsilon J(1-\cos\theta)\defi E~,
\label{e.constraint_CC}
\end{equation}
which is a perfectly meaningful classical condition.
In fact, reminding that 
$\omega M|\alpha|^2=\exval{\alpha|\hat{H}_\smg|\alpha}:=E_\smg(|\alpha|)$ 
and $\epsilon J(1-\cos\theta)=
\exval{\Omega|\hat{H}_C|\Omega}:=E(\theta)$, 
Eq.~\eqref{e.constraint_CC} tells us that the classical state identified 
by the representative point $(q, p)$ in the phase-space of $\Gamma$ (see below for the definition
of $(q,p)$)
is accessible to the system $\Gamma$ 
at time $t$, as marked by the clock $C$ in the state with 
representative point $(\theta,\phi)$, only if $\Gamma$ and $C$ have the 
same energy (we remind that the PaW constraint \eqref{e.Ham_PaW} sets 
the energy of the whole system, but different combinations of subsystems 
energies are still allowed).


Indeed, as shown in Ref.~\cite{FCBCVnat21}, the constraint in 
Eq.~\eqref{e.constraint_CC} defines a map, that leads to the Hamilton 
EoM, between points in the $\Gamma$ and $C$ respective phase spaces (the plane and the 
sphere in this example):
\begin{equation}\label{e. map F}
    \alpha=\sqrt{\frac{E}{M\omega}}\,e^{-i(\eta\,t+\varphi_0)}~,
\end{equation}
where $\eta,\varphi_0\in\mathbb{R}$ and we used the coordinates $(E,t)\in S^2$ for identifying points on the sphere, 
with $E=E(\theta)$ as defined under Eq. \eqref{e.Etheta} and $t=t(\varphi)=\varphi/\epsilon$. In order to describe the classical configurations, we introduce a pair $(q,p)$ of conjugate coordinates for the oscillator by means of the following Darboux chart on the symplectic manifold $\mathbb{C}$ 
\begin{equation}\label{e. Darboux chart}
    \alpha=\sqrt{\frac{M\omega}{2}}\left(q+i\frac{p}{M\omega}\right)~,
\end{equation}
such that $\{q,p\}_\smg=1$, with $\{\cdot,\cdot\}_\smg$ Poisson brackets for $\Gamma$, obtained starting from the measure $d\mu(\alpha)$ as in Ref. \cite{Perelomov86}. Using the chart \eqref{e. Darboux chart}, Eq. \eqref{e.H_alpha} takes the same form of the Hamiltonian function of a unit mass classical harmonic oscillator with frequency $M\omega$, i.e.
\begin{equation}\label{e. oscillatorH classical}
H_\smg(q,p)=\frac{p^2}{2}+\frac{1}{2}(M\omega)^2 q^2~,
\end{equation}
and the classical configurations, surviving the classical limit according to Eq. \eqref{e.beta 3}, look as
\begin{align}\label{e. survivors}
     &\Bigg(E,t;\;q=\frac{\sqrt{2E}}{M\omega}\cos{\left(\eta\,t+\varphi_0\right)},\nonumber\\
     &\qquad\quad p=-\sqrt{2E}\sin{\left(\eta\,t+\varphi_0\right)}\Bigg)\in S^2\times\mathbb{C}~,
\end{align}
with
\begin{equation}\label{e. classical energy}
    E=H_\smg(q,p)=\omega\left(n+\frac{1}{2}\right)\sim (M\omega) \frac{n}{M} \;\mbox{(for $M\rightarrow\infty$)}
\end{equation}
for any $n$ appearing in Eq. \eqref{e.beta 3}. It is now easy to verify that the configurations \eqref{e. survivors} satisfy 
\begin{equation}
\begin{dcases}
    \{p,H_\smg\}_\smg=\frac{M\omega}{\eta}\frac{dp}{dt}\\
    \{q,H_\smg\}_\smg=\frac{M\omega}{\eta}\frac{dq}{dt}
\end{dcases}
\end{equation}
i.e., the Hamilton equations of motion ruling the classical dynamics of $\Gamma$ where, once the arbitrary constant $\eta$ is set equal to $M\omega$, time is recognized, as for the quantum cases of the previous sections, with the parameter $t=\varphi/\epsilon$ provided by the magnetic clock, and $\mathbb{C}$ plays the role of the $\Gamma$ phase-space. Moreover the clock provides to the oscillator the additional parameter $E$ which identifies the energy of $\Gamma$ which, being $H_\smg$ time independent, consistently is a constant of motion. Notice that, as a consequence of having started from quantum physics, the energy values appear discretized as in \eqref{e. classical energy}, but the difference between any two values consistently tends to $0$ as $M\rightarrow\infty$, i.e. as the $\Gamma$ classical limit is taken.  

Let us finally highlight that, being $S_2$ a symplectic manifold, it is possible, starting from the measure $d\mu(\Omega)$ as in Ref. \cite{Perelomov86}, to introduce also for the clock $C$ Poisson brackets $\{\cdot,\cdot\}_\smc$ which are related to the ones $\{\cdot,\cdot\}_\smg$ for $\Gamma$ via the pullback by the map ~\eqref{e. map F}. In particular it is $\{E,t\}_\smc=1$ i.e. the pair $(E,t)$ realizes a Darboux chart on $S_2$. Therefore the latter can be recognized by the oscillator as a well-defined "energy-time phase-space" provided by its clock. Energy and time reveal thus to be conjugate coordinates, but not for the evolving system phase-space (which instead is $\mathbb{C}$).\\

\begin{figure}
    \includegraphics[scale=0.55]{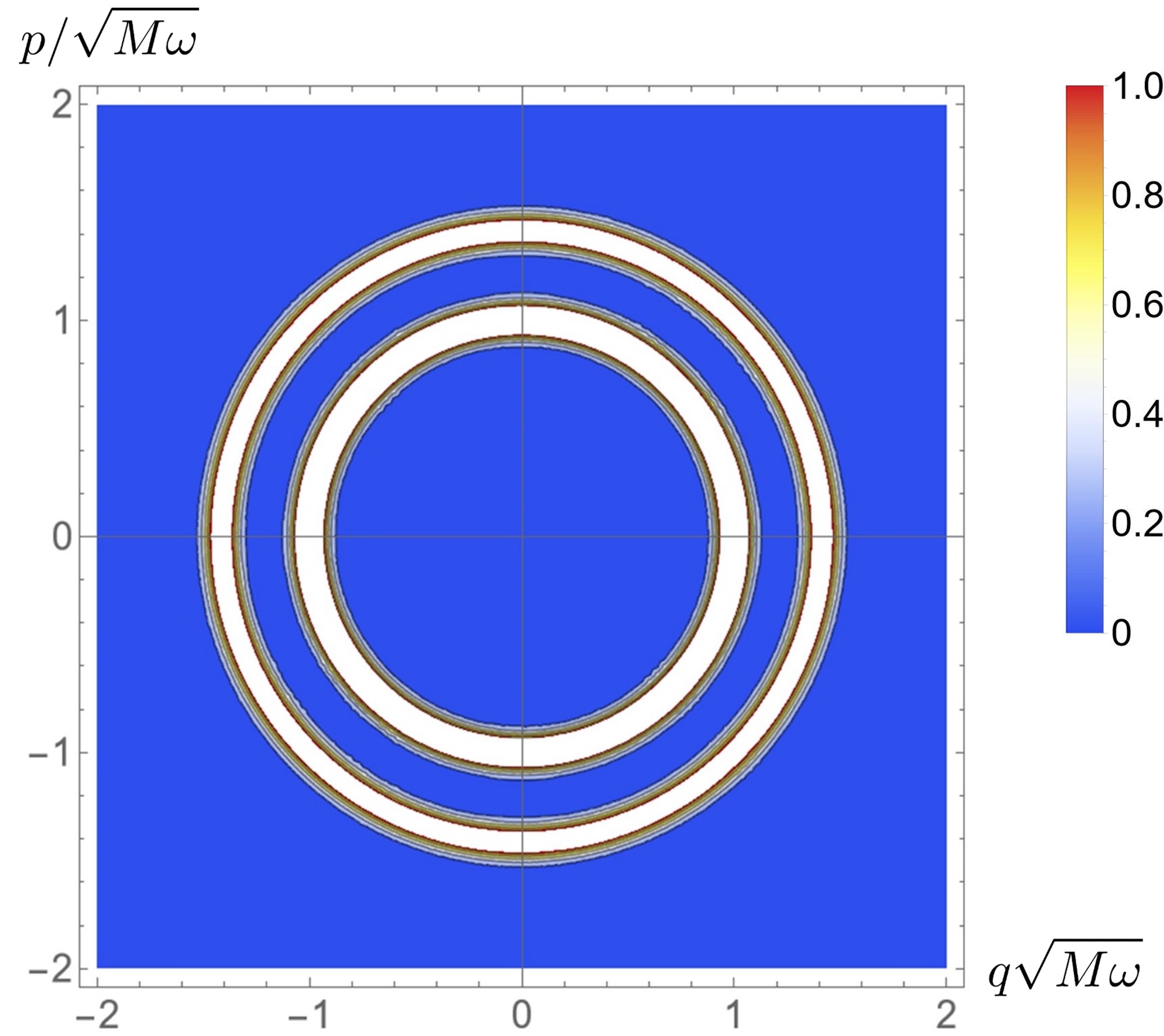} 
    \caption{The contour plot of the marginal probability distribution related to $|\beta|^2$ w.r.t the evolving system phase-space coordinates for the example with $\kappa=3/4$ introduced in the main text. Considering the dimensionless coordinates defined in the figure, the radii are equal to $\sqrt{2}$ and $1$, for $E=M\omega$ and $E=M\omega/2$, respectively. }
    \label{betamargPQ}
\end{figure}

\begin{figure}
    \includegraphics[scale=0.4]{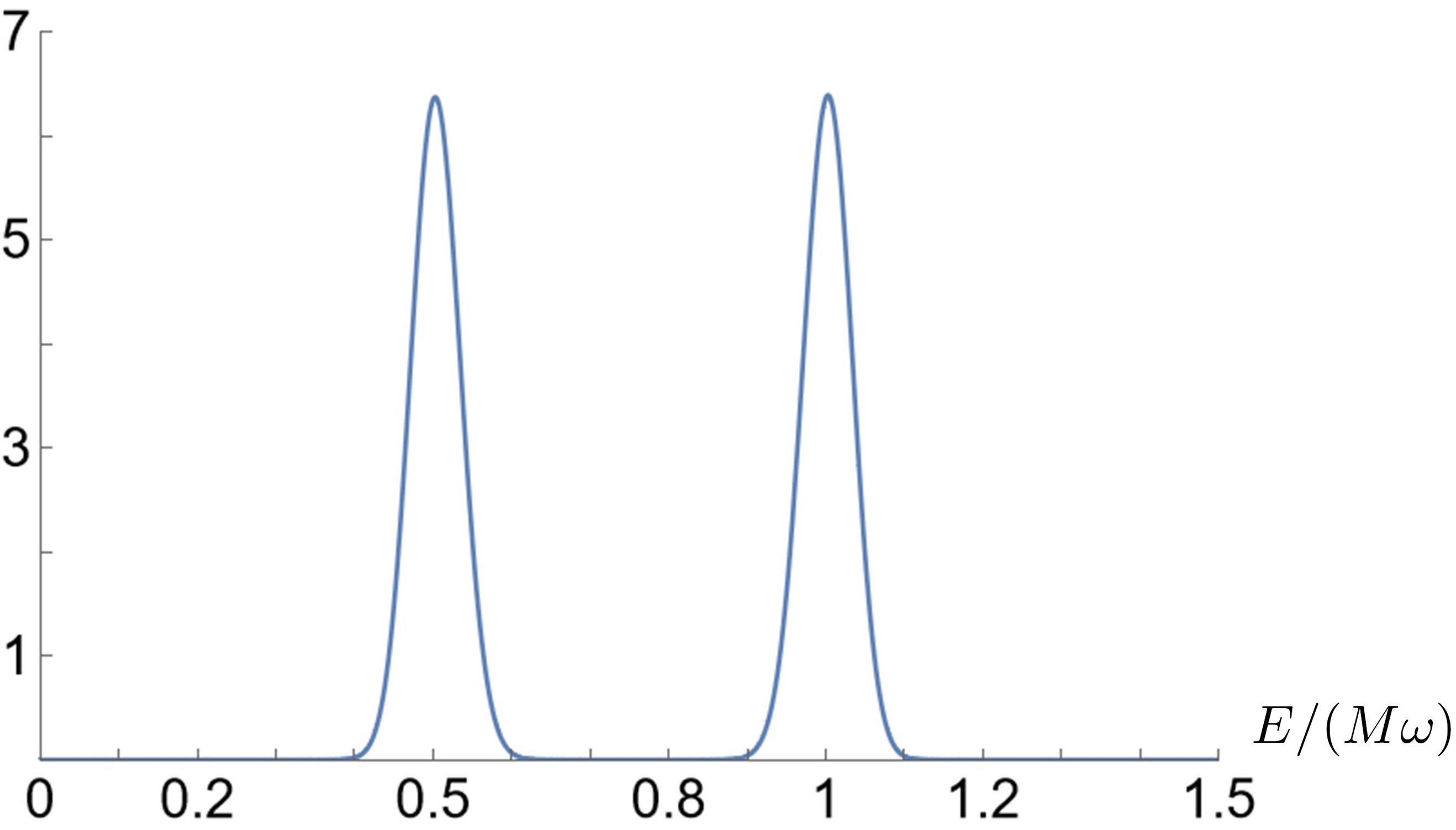} 
    \caption{The section at any constant time of the marginal probability distribution related to $|\beta|^2$ w.r.t the energy-time phase-space for the example with $\kappa=3/4$ introduced in the main text.}
    \label{betamargEt}
\end{figure}

\begin{figure}
\hspace*{-7mm}\includegraphics [scale=0.42]{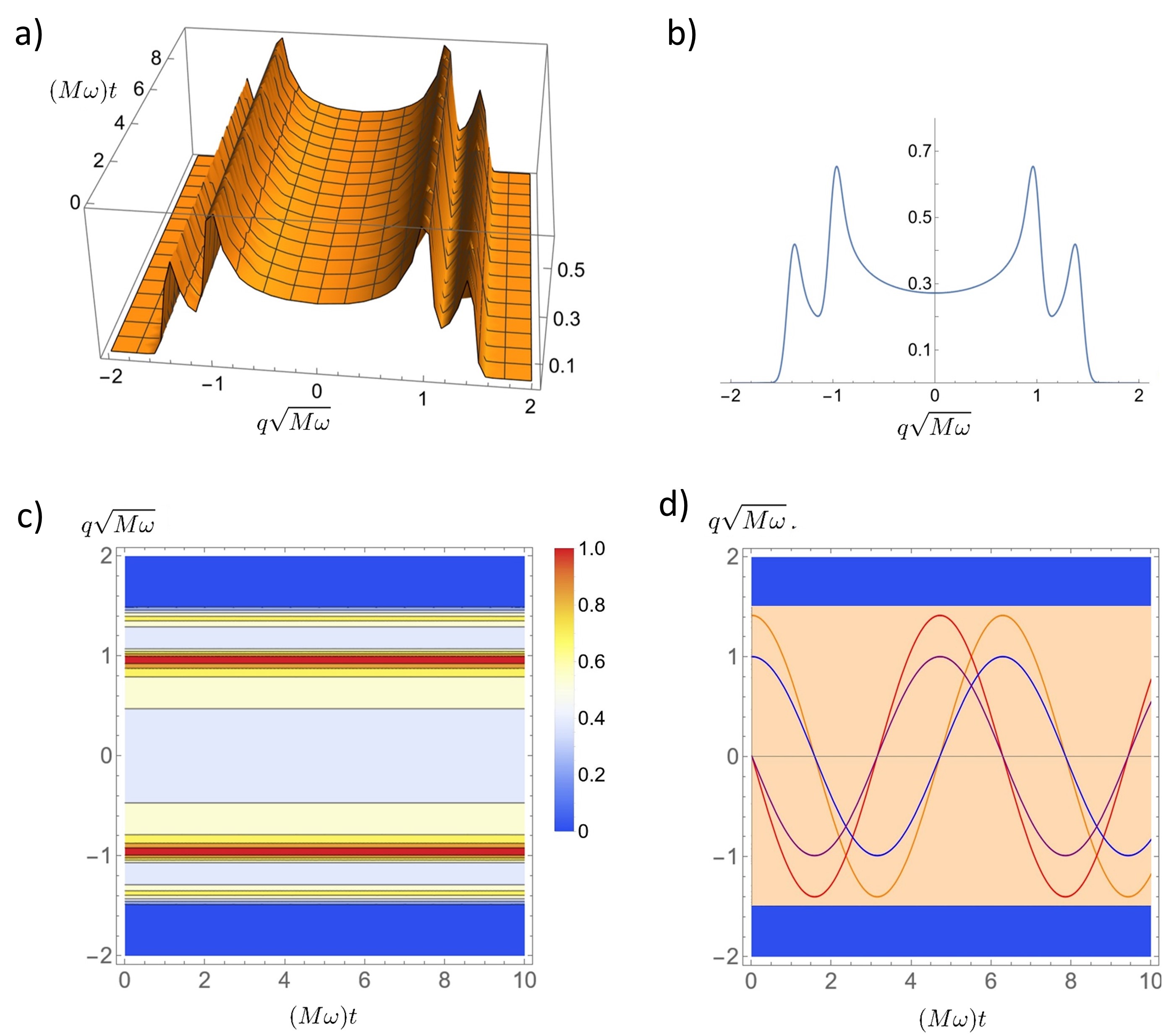}
\caption{a): Graphical representation of the marginal probability distribution related to $|\beta|^2$ w.r.t the space-time coordinates and b) its section at any constant time for the example with $\kappa=3/4$ introduced in the main text. The contour plot of the above distribution is reported in c), while the emergence of the related dynamics, $q$ vs $t$ is reported in d). The light orange region is the support of the marginal probability when $M\rightarrow\infty$.}
\label{betamargQt}
\end{figure}

\noindent In order to verify the consistency of the above construction, we look at the graphical representations of the marginal probability distribution related to $|\beta|^2$ w.r.t. the evolving system phase-space, described by $(q,p)$, w.r.t. the energy-time phase-space, described by $(E,t)$, and w.r.t. the space-time, described by $(q,t)$. General expressions for such distributions are obtained in App. \hyperlink{a. MPD}{B}. We now consider the case $\kappa=3/4$, $r=2/3$, $M=170$ and, in order to have some control over the figures, choose $c_m=(1/\sqrt{2})(\delta_{n_m,M}+\delta_{n_m,M/2})$ allowing only the two energy values $E=M\omega$ and $E=M\omega/2$. We start from Fig.~\ref{betamargPQ} for the evolving system phase-space: the configurations consistently lie in two circumferences whose radii are fixed by the allowed energy values.
\noindent As it concerns the energy-time phase-space, the marginal probability distribution in Fig.~\ref{betamargEt} is, as expected, peaked at any time on $E=M\omega$ and $E=M\omega/2$.
\noindent Let us now consider the space-time. It is possible to show, via the properties of the Dirac delta function appearing for $M\rightarrow\infty$ (see App. \hyperlink{a. MPD}{B}), that, as in the panels a) and b) of Fig.~\ref{betamargQt}, the marginal distribution is proportional to
$\frac{\Theta(2-Q^2)}{\sqrt{(2-Q^2)}}+\frac{\Theta(1-Q^2)}{\sqrt{(1-Q^2)}}$ for any time value,
with $Q\defi  q\sqrt{M\omega}$ and $\Theta$ the Heaviside step function. Consequently the support represented in panel c) of Fig.~\ref{betamargQt} tends to the region $\mathbb{R}\times[-\sqrt{2},\sqrt{2}]$. Indeed, according the energy values, the maximum value achieved by $Q$ during the time evolution is $\sqrt{2}$. Actually, once the map \eqref{e. map F} is introduced, the dynamics emerging from $H_\smg(q,p)$ takes place just on $\mathbb{R}\times[-\sqrt{2},\sqrt{2}]$. The mentioned dynamics is described by panel d) of Fig.~\ref{betamargQt} for $E=M\omega$ (red and orange lines) and $E=M\omega/2$ (purple and blue lines), with the initial conditions $q(t=0)=0$ or $q(t=0)=q_{max}$. The latter is the maximum value achieved by the position $q$, therefore $q_{max}=\sqrt{2/(M\omega)}$ for $E=M\omega$ and $q_{max}=\sqrt{1/(M\omega)}$ for $E=M\omega/2$.\\

\begin{figure}
\includegraphics [width=\linewidth]{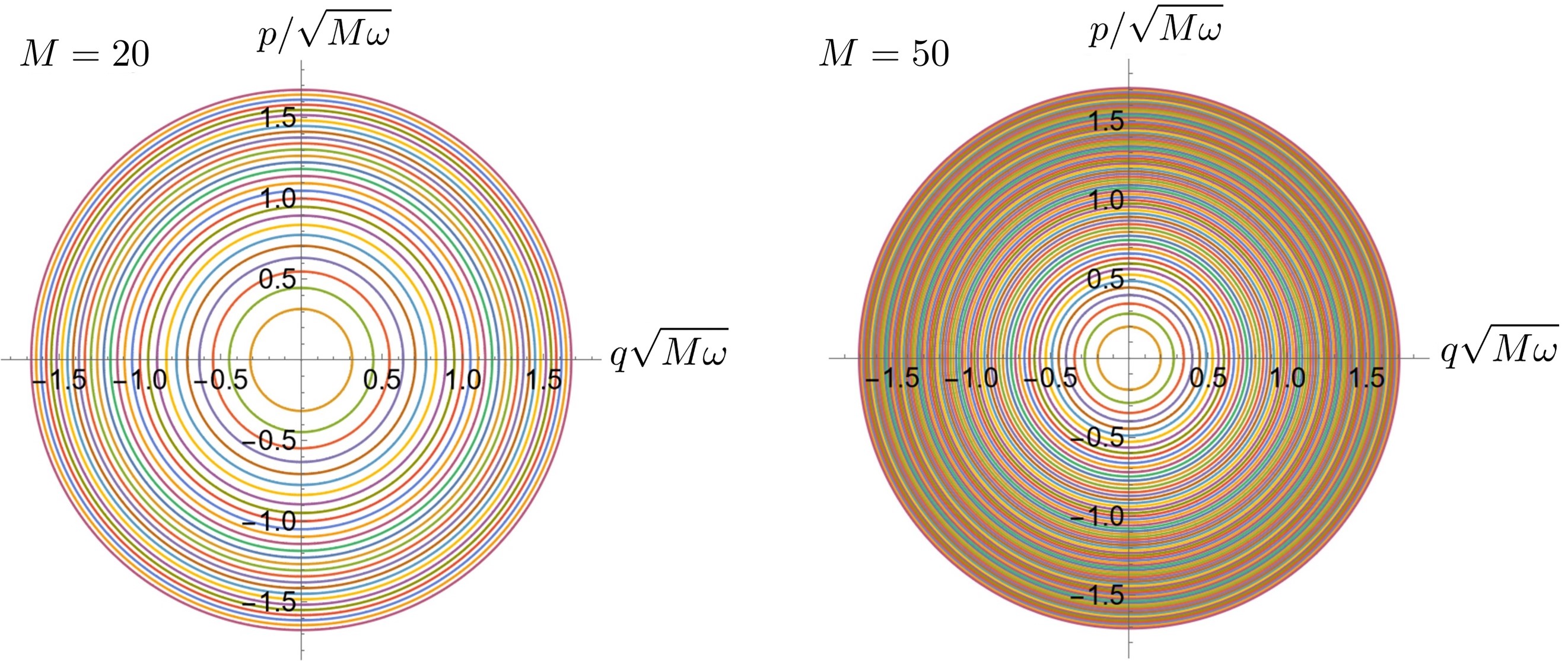}
\caption{The evolving system phase-space with the admitted orbits when $c_m\neq 0$ $\forall m$ and $\kappa=3/4$, $r=2/3$ for $M=20$ (on the left) or $M=50$ (on the right). Using the units in the figure, the orbits are concentric circles with radii $\sqrt{2n/M}$, $n\in\mathbb{N}$. If $M\rightarrow \infty$, the radii are bounded between $0$ and $\sqrt{4\kappa}$.}
\label{dense PQ phase space}
\end{figure}

\begin{figure}
\includegraphics [width=\linewidth]{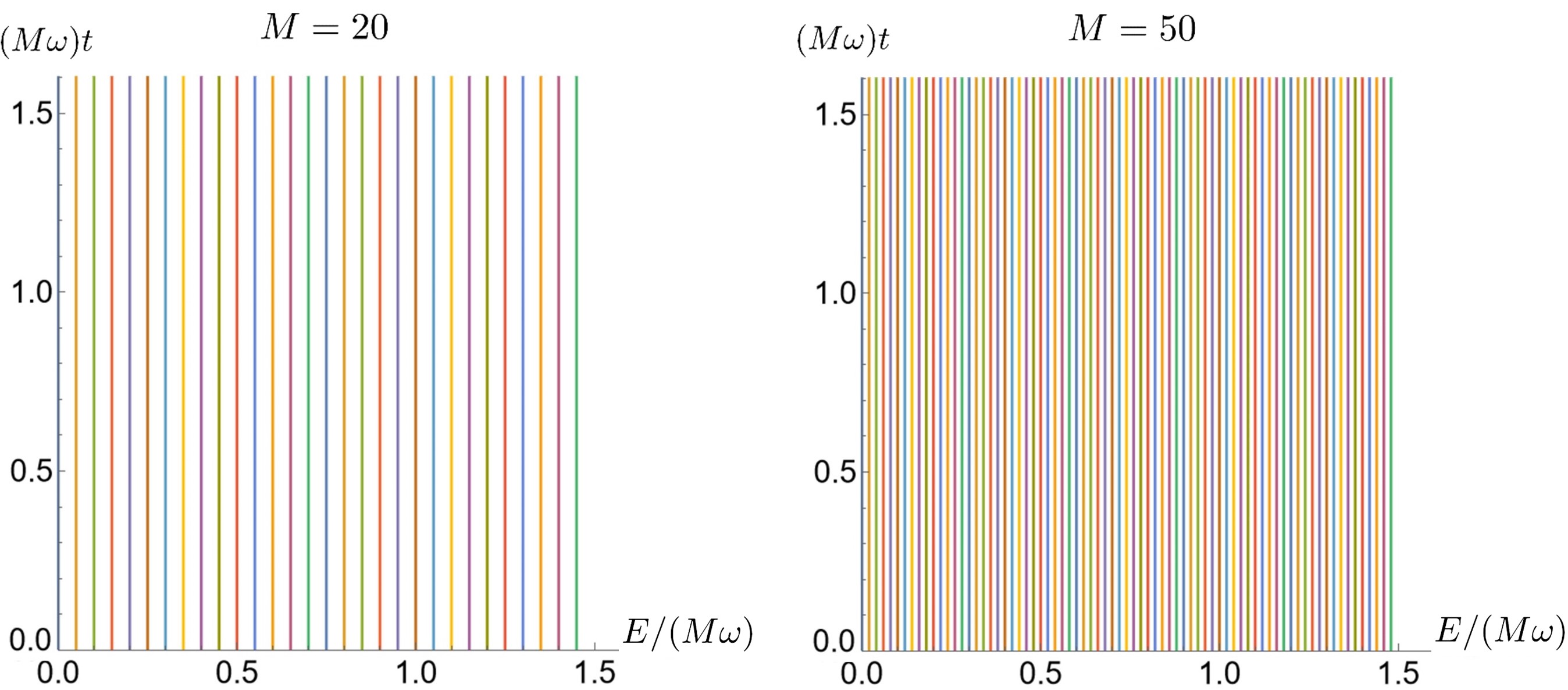}
\caption{The energy-time phase-space with the admitted orbits when $c_m\neq 0$ $\forall m$ and $\kappa=3/4$, $r=2/3$ for $M=20$ (on the left) and $M=50$ (on the right). The orbits are vertical lines satisfying Eq.~\eqref{e. classical energy} for $n\in \mathbb{N}$. If $M\rightarrow \infty$, the lines densely populate the region $0<E/(M\omega)<3/2$.
}
\label{dense Et phase space}
\end{figure}

Let us finally study the case $\kappa=3/4$, $r=2/3$ with every possible state appearing in the decomposition \eqref{e.para_Dicke}. In other words, we assume $c_m\neq 0$ for any $m\in\mathcal{A}$ such that $\dket{\Psi}$ is entangled. As we know, the distance between two subsequent energy values tends to $0$ when $M\rightarrow\infty$, therefore the evolving system phase-space and the energy-time one are covered more and more densely by the admitted orbits, as depicted in Figs.~\ref{dense PQ phase space}-\ref{dense Et phase space}, in accordance with classical physics.
\noindent Notice that the phase-spaces are bounded by the maximum energy value $3M\omega/2$. This constraint may sound puzzling as we are used to consider clocks as
objects whose capability of marking time is not limited by the evolving system energy. On the other hand it easy to show via Eqs.~\eqref{e.constraint_QQ} and \eqref{e. classical energy} that $\kappa\gg 1$, i.e. $J\epsilon\gg M\omega$, is a necessary condition to allow $\Gamma$ to wander in the entire $\mathbb{C}$. The meaning of the above requirement was introduced in the previous section: in order for $C$ to properly work as a clock for $\Gamma$, it must be characterized by a much bigger energy-scale than that of the evolving system, no matter whether this latter is described by a quantum or a classical theory.

\section*{DISCUSSION AND CONCLUSIONS}
\label{s.conclusions}

The actual realization of the PaW mechanism we have considered in this paper
allows us to bring to light, and improve our understanding of, many facets of the
mechanism leading to the definition of time starting from a fully quantum 
description. In contrast with other implementations of the PaW mechanism 
discussed in the literature, which consider continuous and unbounded spectra
for the clock, and possibly for the system too, our construction is built upon 
two paradigmatic quantum systems with a discrete spectrum and, for the spin 
system, even with a separable Hilbert space. As a consequence, as we have shown 
in full detail in Sec.~\ref{s.QQ}, the implementation of the PaW constraint 
\eqref{e.PaW_stationary} leads to strict conditions on the parameters setting 
the energy scales of the two systems and on the allowed states appearing in the
global entangled state $\dket{\Psi}$, which may limit the capability of the clock
to describe the dynamics of the evolving system $\Gamma$ to a rather small subset of 
its Hilbert space. However, as discussed in Sec.~\ref{s.CQ}, such limitations are 
removed, and the possibility to describe the full dynamics of $\Gamma$ is recovered,
when the macroscopic (classical limit) of the clock-system is taken in order to 
reconcile the quantum definition of time by the PaW mechanism with the usual 
classical "time" variable appearing in the Schr\"odinger equation. The steps 
followed in Sec.~\ref{s.CQ} showed that taking the macroscopic limit of the clock
$J\to\infty$ is not by itself enough to reach the goal: indeed, this must be
accompanied by a proper choice of the energy scale of the clock, that must be 
large enough to allow the dynamical system to explore its full Hilbert space. 
The specific conditions expressed in Eqs.~\eqref{e.constraint_QQ}-\eqref{e.J_constr},
and further specialized in the paragraph around Eq.~\eqref{e.constraint_CQ}, when the 
classical limit of the clock is taken, clearly depends on the algebras of the 
actual models describing the evolving system $\Gamma$ and the clock $C$, but they
allowed us to illustrate by a specific example the general property, implied by 
the PaW mechanism, that it is only the combined effect of the macroscopic limit 
and of the magnitude 
of the energy of the clock, that allows us to recover the continuous parameter
traditionally employed to follow the evolution of a quantum system $\Gamma$.
In the same section, Eqs.~\eqref{e.Etheta}-\eqref{e.stat_Schroedinger}, we 
were also able to show how the parameters defining the
GCS state of the clock provide not only the description of the dynamics, but
also define the energy of the evolving system, proving once again that the 
relationship between energy and time originates from the clock, as we already 
observed in Ref.~\cite{FCBCVnat21} while discussing the time-energy
uncertainty relationship within the framework of the PaW mechanism.
The emergence of the classical dynamics when also the evolving system 
becomes macroscopic, discussed in the last section, also benefits from
the possibility given by an actual example of implementation of the PaW
mechanism: Indeed, in addition to the definition of proper conjugate 
coordinates for the evolving system, obeying to the Hamilton's classical
equation of motion, already obtained in Ref.~\cite{FCBCVnat21}, we had the
opportunity to show how the marginal probability distribution defined 
within the fully quantum framework, naturally blends into the expected 
classical space-time and phase-space orbits of the dynamical system in 
the macroscopic, classical limit. 

\section*{ACKNOWLEDGMENTS}
\label{s. Acknowledgements}

\noindent The authors acknowledge financial support from PNRR 
MUR project PE0000023-NQSTI funded by European Union—NextGenerationEU. The 
authors thank Caterina Foti and Nicola Pranzini for useful discussions. This work 
is done in the framework of the Convenzione operativa between the Institute 
for Complex Systems of the Consiglio Nazionale delle Ricerche (Italy) and 
the Physics and Astronomy Department of the University of Florence.
\\

\hypertarget{a.A}{\section*{APPENDIX A}}
The conditions \eqref{e.Kappa_constr}-\eqref{e.J_constr} are obtained starting from Eq.~\eqref{e.constraint_QQ} which can be rewritten as $\kappa r=(2n+1)/(2(m+J))$. Therefore, being $n$ and $(m+J)$ natural numbers, it is necessary that, when reduced to the lowest terms, $\kappa r=(2i_n+1)/(2i_m)$ with $i_n,\,i_m\,\in\mathbb{N}$. Moreover, since $2n+1=2(m+J)u=(2i_n+1)(m+J)/i_m$ has to be an odd number, it must be $m+J=i_m(2l+1)$ with $l\in\mathbb{N}$. More precisely, due to the constraint $m+J<2J$, it is $l\leq\lfloor J/i_m-1/2\rfloor$ so that $J\geq 3i_m/2$ in order to ensure that at least two values of $m$ are allowed. Finally, by means of Eq.~\eqref{e.constraint_QQ}, we obtain $n=i_n(2l+1)+l$.

\hypertarget{a. MPD}{\section*{APPENDIX B}}
In this appendix we obtain the Marginal Probability Distributions (MPDs) related to $|\beta|^2$ represented in the Figs.~\ref{betamargPQ}-\ref{betamargEt}-\ref{betamargQt} of Sec.~\ref{s.CC}.

\section*{B.1.$\;\;$ MPD w.r.t. the evolving system phase-space coordinates}
The MPD related to $|\beta|^2$ w.r.t. the evolving system phase-space, represented in Fig.~\ref{betamargPQ}, is obtained from Eq.~\eqref{e. beta 2} by means of the resolution of identity given by the GCS on $\mathcal{H}_\smc$, from which
\begin{equation}\label{C1}
\int_{S_2}d\mu(\Omega)|\beta|^2=\sum_{m\in\mathcal{A}}|c_m|^2\left|\braket{\alpha|n_m}\right|^2~.
\end{equation}
Considering now the expression of $\braket{\alpha|n}$ in Ref.~\cite{Perelomov86}, the measure $d\mu(\alpha)$ below Eq.~\eqref{e. double para} and the chart \eqref{e. Darboux chart} on $\mathbb{C}$, Eq.~\eqref{C1} reads
\begin{align}
&\frac{M}{2\pi}\sum_{m\in\mathcal{A}}|c_m|^2 \,e^{-\frac{M}{2}\left(Q^2+P^2\right)}\,\left(\frac{M}{2}(Q^2+P^2)\right)^{n_m}\frac{1}{(n_m)!} \nonumber\\
&\xrightarrow{M\rightarrow \infty}\frac{1}{\pi}\sum_{m\in\mathcal{A}}|c_m|^2\,\delta\!\left(Q^2+P^2-2n_m/M\right)~,
\end{align}
with  $Q\defi  q\sqrt{M\omega}$ and $P\defi  p/\sqrt{M\omega}$. Finally, for $M\rightarrow \infty$, the configurations lie in circumferences with radii $2n/M=E/(M\omega)$, where the last equality follows from Eq.~\eqref{e. classical energy}.

\section*{B.2.$\;\;$ MPD w.r.t. energy-time  variables}

The MPD related to $|\beta|^2$ w.r.t. the energy-time phase-space, represented in Fig.~\ref{betamargEt}, is obtained from Eq.~\eqref{e. beta 2} by means of the resolution of identity given by the GCS on $\mathcal{H}_\smg$, from which
\begin{equation}\label{C2}
    \int_{\mathbb{C}}d\mu(\alpha)|\beta|^2=\sum_{m\in\mathcal{A}}|c_m|^2\left|\braket{\Omega|J,m}\right|^2=\chi^2(\theta)~,
\end{equation}
as defined in Eq.~\eqref{e.chi2_def}. Therefore, defining $\mathtt{e}\defi  E/(M\omega)$, considering Eqs.~\eqref{e.chi2} and \eqref{e.Etheta} together with the measure $d\mu(\Omega)$ below \eqref{e.para_Psi}, Eq.~\eqref{C2} reads
\begin{align}
\chi^2(\mathtt{e})&=\frac{2J+1}{2\kappa}\sum_{m\in\mathcal{A}}|c_m|^2 \binom{2J}{m+J}\left(1-\frac{\mathtt{e}}{2\kappa}\right)^{J-m}\left(\frac{\mathtt{e}}{2\kappa}\right)^{J+m}\nonumber\\
&\xrightarrow{J\rightarrow \infty} \sum_{m\in\mathcal{A}}|c_m|^2\delta\left(\mathtt{e}-\kappa(1+m/J)\right)\nonumber\\
&=\sum_{m\in\mathcal{A}}|c_m|^2\delta\left(\mathtt{e}-n_m/M\right)~,
\end{align}
where we used Eq.~\eqref{e.constraint_QQ}.

\section*{B.3.$\;\;$ MPD w.r.t. the space-time coordinates}
The MPD related to $|\beta|^2$ w.r.t. the space-time is
\begin{equation}\label{C3}
\left(\frac{2J+1}{2\kappa}\int_0^{3/2} d\mathtt{e}\right)\;\left(\frac{M}{2\pi}\int_{-\infty}^{+\infty}dP\right)\,|\beta|^2    
\end{equation}
with $\mathtt{e}\defi  E/(M\omega)$ and $P\defi  p/\sqrt{M\omega}$. In order to have some control over the calculation, let us assume that, as in the example of the main text, $\mathrm{dim}(\mathcal{H}_\smg)=2$, i.e. that only two couples $(m_i,n_i\defi n_{m_i})$ $i=1,2$ are allowed by Eq.~\eqref{e. beta 2}. Moreover let us write $\beta=c_1z_1+c_2z_2$ where $c_i\defi  c_{m_i}$, $z_i\defi  z_i^\smc z_i^\smg$ and $z_i^\smc\defi  \braket{\Omega|J,m_i}$, $z_i^\smg\defi  \braket{\alpha|n_i}$. Once implemented, the changes of coordinates $(\theta,\varphi)\Rightarrow(\mathtt{e},t)$ for $S_2$ and $\alpha\Rightarrow (P,Q)$ for $\mathbb{C}$ with $t\defi \varphi/\epsilon$ and $Q\defi  q\sqrt{M\omega}$, we decompose Eq.~\eqref{C3} as the sum of three integrals $I_1+I_2+I_{int}$ defined via
\begin{equation}
    \!\begin{cases}
    I_i\defi  |c_i|^2\;\cdot\;\left(\frac{2J+1}{2\kappa}\int_0^{3/2} d\mathtt{e}\,|z_i^\smc|^2\right)\;\cdot\;\left(\frac{M}{2\pi}\int_{-\infty}^{+\infty}dP\,|z_i^\smg|^2\right)\vspace*{1mm}\\
    \,\,\quad=|c_i|^2\;\cdot\;\frac{M}{2\pi}\int_{-\infty}^{+\infty}dP\,|z_i^\smg|^2\\
    \\
    I_{int}\defi  2|c_1|\,|c_2|\;\cdot\;\left(\frac{2J+1}{2\kappa}\int_0^{3/2}d\mathtt{e}\,|z_1^\smc|\,|z_2^\smc|\right)\;\cdot\\
    \qquad\quad\cdot\left(\frac{M}{2\pi}\int_{-\infty}^{+\infty}dP\,|z_1^\smg|\,|z_2^\smg|\,\cos(z_{int})\right)
    \end{cases}
\end{equation}
where $z_{int}\defi  (n_1-n_2)\arctan(P,Q)+(m_1-m_2)\varphi+\arg(c_1)-\arg(c_2)$. Considering now the limits $J,M\rightarrow\infty$, $I_i$ can be evaluated via
\begin{align}
    &I_i\xrightarrow{N\rightarrow \infty}\frac{|c_i|^2}{\pi}\;\cdot\;\int_{-\infty}^{+\infty}dP\,\delta\!\left(Q^2+P^2-2n_i/M\right)\nonumber\\
       &=\frac{2|c_i|^2\;\Theta\left(2n_i/M-Q^2\right)}{\pi}\;\cdot\;\int_0^{+\infty}\frac{dP}{|2P|}\nonumber\\
       &\left[\delta\left(P-\sqrt{2n_i/M-Q^2}\right)+\delta\left(P+\sqrt{2n_i/M-Q^2}\right)\right]\nonumber\\
       &=|c_i|^2\;\frac{\Theta\left(2n_i/M-Q^2\right)}{\pi\sqrt{2n_i/M-Q^2}}
\end{align}
with $\Theta$ the Heaviside step function. As concerns $I_{int}$, the integral in $d\mathtt{e}$ can be exactly calculated as
\begin{align}
    &\binom{2J}{m_1+J}^{1/2}\binom{2J}{m_2+J}^{1/2}\binom{2J}{\frac{m_1+m_2}{2}+J}^{-1}\xrightarrow{J\rightarrow \infty}0
\end{align}
for $m_1\neq m_2$, instead the integral in $dP$ can be approximated for $M\rightarrow\infty$ as
\begin{align}
    &\frac{\big((n_1+n_2)/2\big)!}{\sqrt{n_1!n_2!}}\,\cdot\,\frac{\Theta\left((n_1+n_2)/M-Q^2\right)}{\pi\sqrt{(n_1+n_2)/M-Q^2}}\,\cdot\nonumber\\
    &\cdot\left(\cos[...]+\cos[...]\right)\xrightarrow{M\rightarrow \infty}0
\end{align}
for $n_1\neq n_2$, with the factorial extended using the Euler Gamma-function when necessary and where, according to Eqs.~\eqref{e.constraint_QQ}-\eqref{e.Kappa_r_def}, we assumed $n_1,n_2 \propto M$ and $m_1,m_2 \propto J$. Finally, in the classical limit, $I_{int}\rightarrow 0$ and Eq.~\eqref{C3} reads
\begin{equation}
\sum_{m\in\mathcal{A}}|c_m|^2\,\frac{\Theta\left(2n_m/M-Q^2\right)}{\pi\sqrt{2n_m/M-Q^2}}~.
\end{equation}

\bibliographystyle{unsrt}

\end{document}